\pdfoutput=1
\documentclass[10pt]{article}
\usepackage[utf8]{inputenc}
\usepackage[style=base]{caption}
\usepackage{array}
\usepackage{amsmath}
\usepackage{graphicx}

\makeatletter

\providecommand{\tabularnewline}{\\}

\usepackage{mathrsfs}
\usepackage[OE]{express}
\usepackage{braket}

\makeatother

\begin{document}

\title{Implementation of nearly arbitrary spatially-varying polarization transformations:
a non-diffractive and non-interferometric approach using spatial light
modulators}

\author{M.T. Runyon\authormark{1}, C.H. Nacke\authormark{1}, A. Sit\authormark{1},
M. Granados-Baez\authormark{2}, L. Giner\authormark{1}, and
J.S. Lundeen\authormark{1}}

\address{\authormark{1}Department of Physics, Centre for Research in Photonics, University of Ottawa, 25 Templeton Street, Ottawa, Ontario K1N 6N5, Canada}
\address{\authormark{2}Department of Physics, University of Rochester, 275 Hutchison Rd, Rochester, NY 14627, U.S.A.}

\email{\authormark{*}mtrunyon91@gmail.com} 


\begin{abstract}
A fast and automated scheme for general polarization transformations
holds great value in adaptive optics, quantum information, and virtually
all applications involving light-matter and light-light interactions.
We present an experiment that uses a liquid crystal on silicon spatial
light modulator (LCOS-SLM) to perform polarization transformations
on a light field. We experimentally demonstrate the point-by-point
conversion of uniformly polarized light fields across the wave front
to realize arbitrary, spatially varying polarization states. Additionally,
we demonstrate that a light field with an arbitrary spatially varying
polarization can be transformed to a spatially invariant (i.e., uniform)
polarization. 
\end{abstract}



\section{Introduction}

In conventional photonic applications, the polarization state of a
light field is often considered to be uniform across the wave front,
i.e., the same polarization exists at each position in the field's
transverse profile. Conventional optical elements, such as waveplates,
are transversely invariant and, thus, maintain this polarization uniformity.
While these uniform polarization transformations have proven useful
in a variety of quantum and classical applications such as quantum
key distribution (QKD), Stokes polarimetry, and ellipsometry, there
has been considerable recent interest in non-uniform polarization
structures \cite{mitchell:17,salla:17}. In \cite{leuchs:03}, it
was shown that radially polarized light fields can be focused to a
smaller spot than similar uniformly polarized fields. In telecommunications,
spatially distinct regions of the light field can function as independent,
polarization-encoded channels. Non-uniform polarization states push
the boundaries of polarization imaging \cite{zhao:16}, where graphical
information is encoded in a light field in such a way that is invisible
to the naked eye. These nascent applications show that a means of
arbitrarily manipulating polarization in full generality has considerable
potential in both science and industry. This work demonstrates a method
for such manipulation that uses liquid crystal spatial light modulators
(SLMs).

Over the past three decades, SLMs have been used to generate light
fields with specific intensity and phase profiles \cite{clark:16,bolduc:13,arrizon:07}.
Beginning with some initial work in 2000, a few research groups have
also begun producing spatially varying polarization profiles using
SLMs. However, the full potential of this approach has yet to be realized.
In \cite{moreno:12,chen:11}, the set of produced polarization states
were limited to those in a plane of the Poincaré sphere. Other previous
experiments often suffer from an inherent optical loss due to their
diffractive or interferometric design \cite{neil:02,maurer:07,wang:07,franke-arnold:07,han:15,waller:13,maluenda:13,guo:14,chen:15}.
Unlike in classical imaging or telecommunications, where substantial
optical loss can be tolerated, even small loss can stop quantum information
and quantum metrology applications from functioning at all. Moreover,
quantum applications often require sophisticated state transformations,
whereas past work has focused on producing particular non-uniform
polarization states \cite{davis:2000,eriksen:01,kenny:12,estevez:15,zheng:15,galvez:12}.

In our demonstration, SLMs are used without any diffractive or interferometric
methods to exploit the device's ability to perform various polarization
transformations. Following the proposal in \cite{sit:17}, we show
that two sequential SLM incidences will induce a controllable, near- arbitrary 
conversion of the polarization point-by-point across the wave front of a light field.
In our demonstration, both incidences are on the same SLM device.
While routing the light for this scheme necessitates considerable
loss, the setup serves as a proof-of-principle demonstration for a
design that has no inherent loss. Specifically, a setup based on two
transmissive SLMs will only suffer from loss due to technical issues,
such as the SLM array fill-factor or anti-reflection coating performance,
both of which can be improved through technical refinements. In summary,
we demonstrate the production of sophisticated spatially varying polarization
profiles, and, more generally, we show that we can transform
polarizations in a similarly sophisticated manner.

We begin in Section \ref{sec:theory} by reviewing the theory behind
our method, as first described in \cite{sit:17}. The experimental
setup is detailed in Section \ref{sec:expsetup}. The key to our successful
demonstration is a careful procedure, described in Section \ref{sec:calibration},
to calibrate the operation of the device. We present the experimental
results in Section \ref{sec:results}. There, we demonstrate the production
of arbitrary, spatially varying polarization states from a known uniform
polarization state. The degree of control and sophistication that
we can achieve in our scheme is demonstrated by our ability to paint
with polarization: we render Van Gogh's painting \emph{Starry Night}
in elliptical polarization states. We then demonstrate the scheme's
ability to convert arbitrary input polarizations by homogenizing a
beam with dramatic spatial polarization variation. In short, we `heal'
the polarization of an aberrated light field.

\section{Theory}

\label{sec:theory}

\subsection{Polarization transformations with SLMs}

In this work, we restrict ourselves to perfectly polarized fields.
These are represented by reduced Stokes vectors $\mathbf{S}=[s_{1},s_{2},s_{3}]$
in the Poincaré sphere (having orthogonal axes $\mathbf{S_{1}},\mathbf{S_{2}},$
and $\mathbf{S_{3}}$) with $\mathbf{\left|S\right|}^{2}=s_{1}^{2}+s_{2}^{2}+s_{3}^{2}=1$.
The conventions and notation used for this formalism are identical
to those used by Sit \emph{et al.} and can be found in Appendix A
of their work \cite{sit:17}. A general unitary (i.e., lossless and
noiseless) polarization transformation is most simply described by
a rotation $\hat{R}(\zeta,\mathbf{k})$ of a Stokes vector $\mathbf{S}$
about an axis $\mathbf{k}$ of the Poincaré sphere by an angle $\zeta$
known as the retardance.

The work horse behind polarization transformations in our experiment
is the SLM \textendash{} a spatial array of liquid crystal (LC) cells
with a common optical axis. Under standard SLM mounting, this axis
is oriented along the horizontal or vertical laboratory direction,
which in our convention is along the $\pm\mathbf{S_{1}}$ polarizations.
It follows that each cell, labelled by $(i,j)$, in the SLM array
induces the rotation $\hat{R}(\zeta_{ij},\mathbf{S_{1}})$, where
$\zeta_{ij}$ is set by the birefringence of the cell. In turn, the
birefringence of each individual cell can be controlled by an applied
voltage. Consequently, distinct transverse positions $(x_{j},y_{i})$
of a light field incident on an SLM can acquire distinct polarizations,
thereby creating a spatially non-uniform polarization.

\subsection{The two-step scheme}

In order to rotate about other axes in the Poincaré sphere without
physically rotating the SLM, the SLM can be sandwiched between matched
sets of waveplates. In effect, these waveplates rotate the polarization
basis in which the SLM acts. For example, if the SLM is preceded by
a half waveplate (HWP) with its optic axis oriented at $22.5^{\circ}$
to the horizontal and followed by another HWP at $22.5^{\circ}+90^{\circ}$,
then it will effectively rotate the polarization about $\mathbf{S_{2}}$,
the diagonal and anti-diagonal polarization axis.

\begin{figure}[h!]
\centering \includegraphics[width=0.5\textwidth]{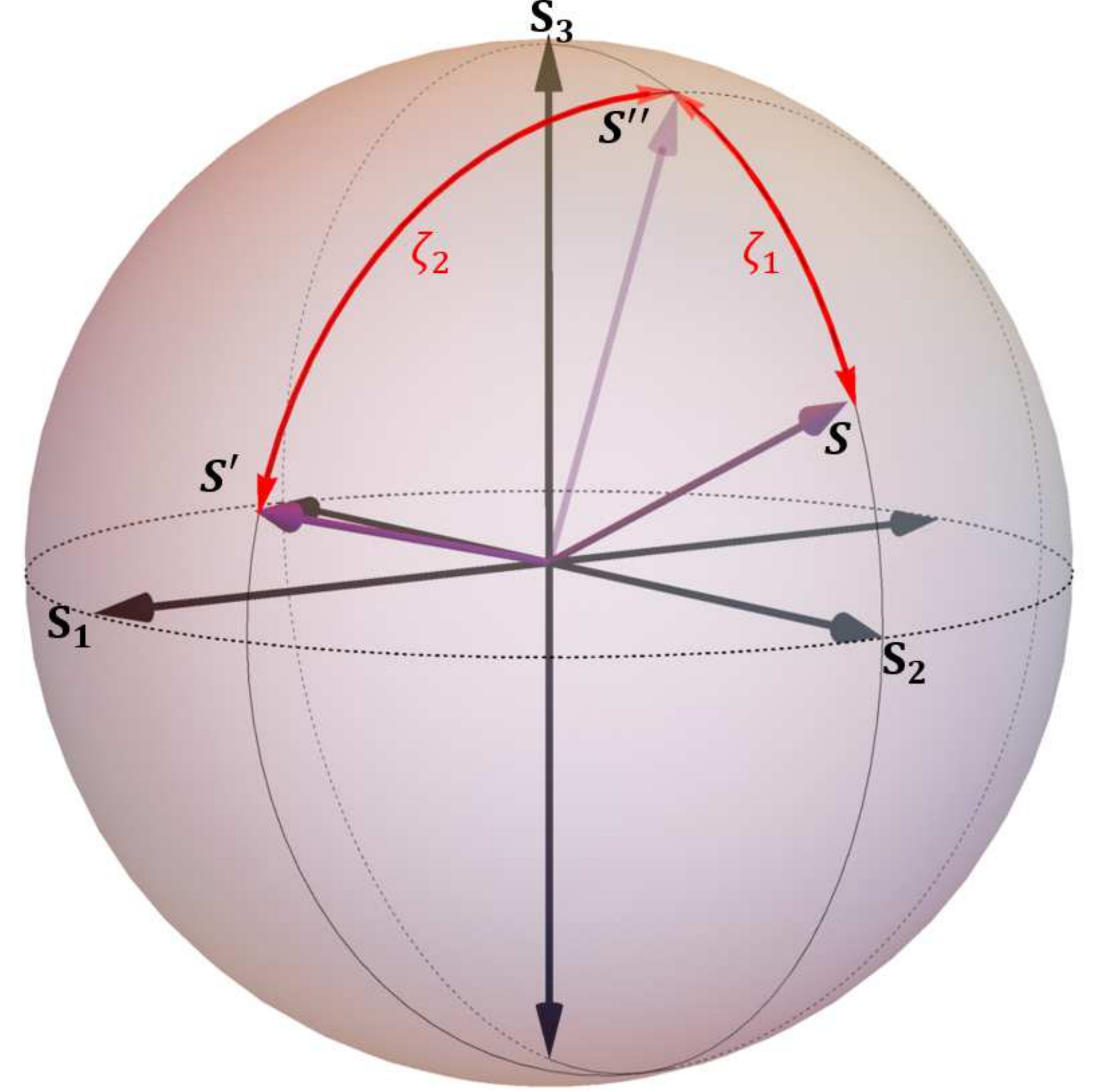}
\captionsetup{width=0.7\textwidth} 
\caption{The two-step scheme. The first step is a rotation in the Poincaré
sphere about $\mathbf{S_{1}}$, the second is a rotation about $\mathbf{S_{2}}$.
The point at which the first rotation ends and second rotation starts
is denoted by $\mathbf{S}"$. This intermediate point is described
by Eq. (\ref{eq:sm}) in the appendix.}
\label{fig:twostep} 
\end{figure}

This second axis comes in to use in the two-step transformation scheme,
theoretically described by Sit \emph{et al.} in \cite{sit:17} and
shown in Fig. \ref{fig:twostep}. This scheme is composed of two successive
rotations with fixed rotation axes $\mathbf{S_{1}}$ and $\mathbf{S_{2}}$,
i.e., $\hat{R}(\zeta_{2,ij},\mathbf{S_{2}})\hat{R}(\zeta_{1,ij},\mathbf{S_{1}})$.
It converts an arbitrary input state $\mathbf{S}{}_{ij}=[s_{1,ij},s_{2,ij},s_{3,ij}]$
to any arbitrary output state $\mathbf{S}'_{ij}=[s'_{1,ij},s'_{2,ij},s'_{3,ij}]$
with $(s'_{2,ij})^{2}\leq1-(s_{1,ij})^{2}$. The initial Stokes vector
is rotated about $\mathbf{S_{1}}$ until the desired $s_{2,ij}$ component
of $\mathbf{S}'_{ij}$ is reached. Then, this intermediate Stokes
vector is rotated about $\mathbf{S_{2}}$ until the desired $s_{1,ij}$
component of $\mathbf{S}'_{ij}$ is reached. The sign of the last
component $s_{3,ij}$ is set by either the first or second rotation
and its magnitude is set by our normalization, $|\mathbf{S}'_{ij}|=1$.
The exact formulae for the rotation angles $\zeta_{1,ij}$ and $\zeta_{2,ij}$
are non-trivial. They are derived in \cite{sit:17} and given in Appendix
\ref{subsec:retardances} for convenience. In our experiment, two
successive incidences on a single SLM compose the two steps in the
two-step scheme.

While this two-step transformation is, seemingly, almost completely
general, it is not. A crucial distinction is that the input and output
states must be known \emph{a priori} since the rotation angles $\zeta_{1,ij}$
and $\zeta_{2,ij}$ depend on them \cite{sit:17}. The output state
can be chosen as a target, whereas the input state can either be determined
experimentally point-by-point or produced via a trusted procedure
(e.g., a polarizer). Consequently, the two-step scheme could not be
used to completely compensate for a general unitary polarization transformation,
such as those occurring in optical fibres. Any chosen pair of orthogonal
states could be compensated, but no others.

Still, the capabilities of the two-step scheme are substantial. For
example, as long as the input state $\mathbf{S}{}_{ij}$ has $s_{1,ij}=0$,
any output polarization $\mathbf{S}'_{ij}$ can be attained. Conversely,
any input polarization state can be transformed to any desired $\mathbf{S}'_{ij}$
that has $s'_{2,ij}=0$.

\subsection{Characterizing spatially varying polarizations}

To operate and evaluate the two-step scheme, we completely determine
the polarization $\mathbf{S}'_{ij}$ point-by-point across the output
field using Stokes polarimetry (see Section \ref{sec:expsetup} for
technical details). In order to visualize the system performance,
we will create two dimensional density plots for each of the three
Stokes components. While these are a complete characterization of
the output field's polarization, it is more useful to distill relevant
characteristics of the $\mathbf{S}{}_{ij}$ matrix into a quantitative
performance metric.

First, we identify a spatial region of interest $A$ in the field
in which we reduce the $\mathbf{S}{}_{ij}$ matrix to an average Stokes
vector given by, 
\begin{equation}
\bar{\mathbf{S}}=\frac{\sum_{i,j=1}^{A}\mathbf{S}{}_{ij}}{|\sum_{i,j=1}^{A}\mathbf{S}{}_{ij}|}.\label{eq:fwtm}
\end{equation}
In most measurements, we will be analyzing the full spatial extent
of the field. However, we limit the region $A$ to a circle with a
diameter equal to the full width at tenth maximum (FWTM) beam waist
in order to avoid the noisy signal in regions of vanishing light intensity.
In some measurements, we will set $A$ to be a wedge (i.e., circular
sector) of this FWTM circle. The wedge is chosen so that it is contained
within a chosen spatial quadrant of the beam.

Our first metric is reminiscent of the well-known degree of polarization
\cite{wolf:99}, but is specifically for non-uniformly polarized light
fields. We define the uniformity $U$ of the polarization of the light
field to be the length of the average Stokes vector, 
\begin{equation}
U\equiv\left|\bar{\mathbf{S}}\right|.\label{eq:uniformity}
\end{equation}
The uniformity metric varies from $U=0$ for spatially varying polarizations
to $U=1$ for a light field that has the same polarization everywhere
in region $A$. For example, if half the region is any particular
polarization and the other half is the orthogonal polarization then
$U=0$.

Our second metric, fidelity $F,$ parametrizes the similarity of two
polarization states, 
\begin{equation}
F\left(\bar{\mathbf{S}}_{\mathrm{exp}},\mathbf{S}_{\mathrm{ideal}}\right)\equiv\frac{1}{2}\left(1+\bar{\mathbf{S}}_{\mathrm{exp}}\cdot\mathbf{S}_{\mathrm{ideal}}\right).\label{eq:fidelity}
\end{equation}
The $\cdot$ symbol denotes the standard vector dot product. The average
Stokes vector $\bar{\mathbf{S}}_{\mathrm{exp}}$, typically calculated
from the output of the experiment, is potentially imperfect in two
ways; the direction of the Stokes vector can differ from $\mathbf{S}_{\mathrm{ideal}}$
and it might have a norm of less than one, signifying non-uniformity.
In contrast, the ideal state $\mathbf{S}_{\mathrm{ideal}}$, typically
the target state, always has a norm of one. This ensures that $F$
matches a well-known measure from quantum information theory known
as the Uhlmann fidelity \cite{josza:94}. The fidelity varies from
$F=0$, for two orthogonal polarization states, to $F=1$, for two
identical polarization states. These two bounds can only be reached
when $\bar{\mathbf{S}}_{\mathrm{exp}}$ is uniform, $U=1$.

With these two metrics in hand, we will be able to characterize how
well we can produce a target state and the degree to which we can
change the spatial variation of polarization across a region $A$.

\section{Experimental Setup}

\label{sec:expsetup} 
\begin{figure}[h!]
\centering \includegraphics[width=1\textwidth]{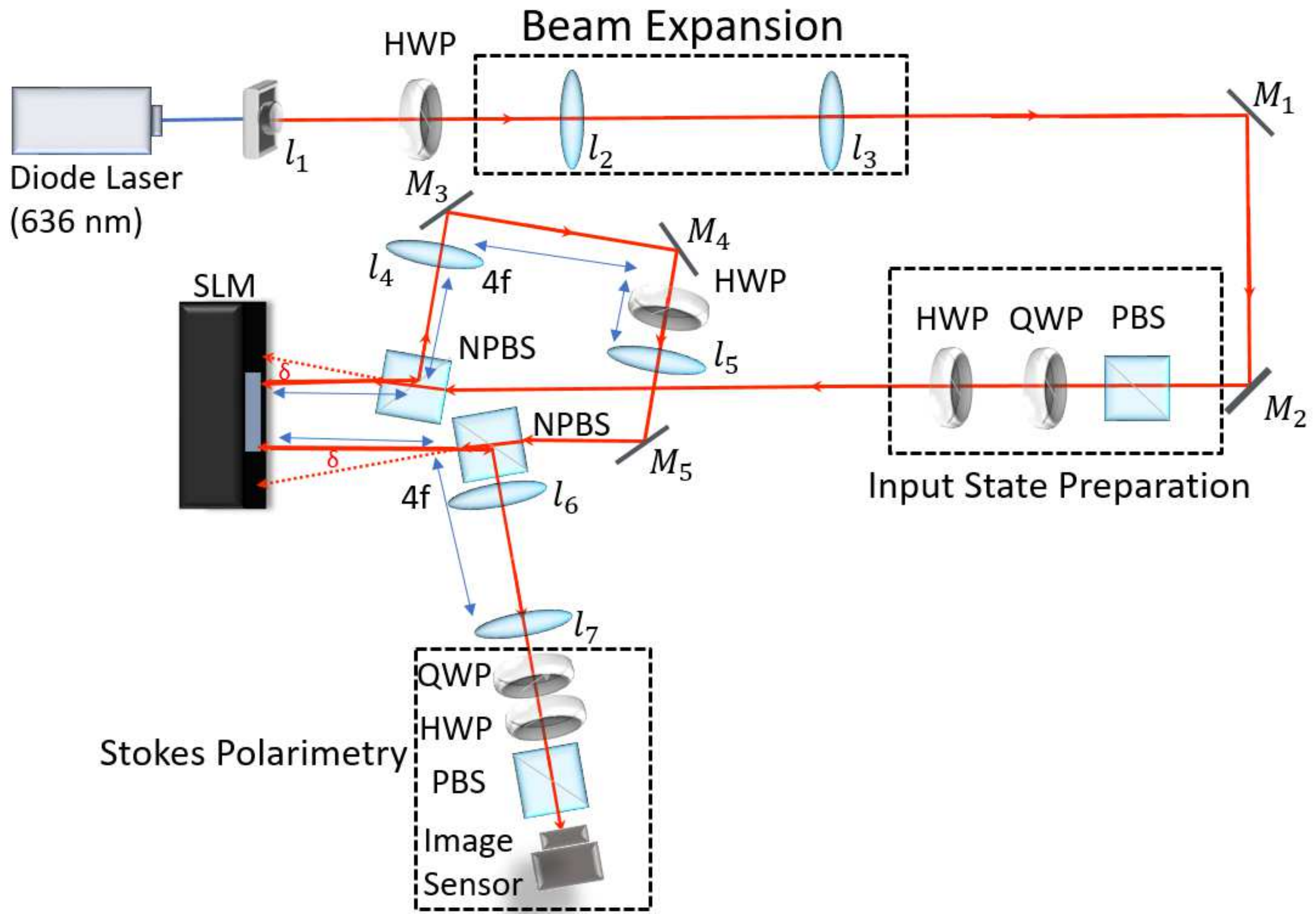} \caption{Schematic of experimental setup (see text for details). The NPBS are
tilted by a small angle $\delta$ to ensure that unwanted reflections
from the faces of the NPBS are directed away. Lenses $l_{4}-l_{7}$
each have a focal length $f=100mm$. Note: $\textrm{M}_{i}$: mirror,
SLM: spatial light modulator, HWP: half waveplate, QWP: quarter waveplate,
NPBS: non-polarizing beamsplitter, PBS: polarizing beamsplitter.}
\label{fig:schematic} 
\end{figure}

Ideally, one would use two transmissive SLMs in sequence to implement
the two polarization rotations of the two-step scheme. The only loss
would be from non-ideal performance of the optical elements. Currently,
however, reflective SLMs have better performance specifications and
are more common in optics laboratories. Consequently, we adapt the
simple (and ideal) transmissive SLM scheme to reflective SLMs by using
50:50 non-polarizing beamsplitters (NPBS) to separate the incident
and reflected light. This has the disadvantage that 50\% of the light
is lost at each encounter with a NPBS. Since each of the two NPBSs
is passed through twice, the total optical transmission is at most
6.25\%. Nonetheless, most of the techniques developed here will be
applicable in the future to the lossless, transmissive SLM scheme.
In this way, it functions as proof-of-principle demonstration and
a prototype. Moreover, using a reflective SLM enables us to use the
two halves of a single SLM array for the two rotations. A schematic
of the setup can be found in Fig. \ref{fig:schematic}.

We begin by describing our source of light and how we produce input
polarizations. A continuous wave single-mode fibre-pigtailed diode
laser is operated at a sub-threshold bias current in order to achieve
a full-width half-maximum spectral bandwidth of 13.2 nm. This is critical
for mitigating interference between reflections from, say, the front
glass surface and the back silicon surface of the SLM, a topic discussed
in \cite{moreno:14}. The fibre output is collimated and then expanded
by lenses $l_{1},$ $l_{2},$ and $l_{3}$ to have a $1/e^{2}$ beam
radius of 1.28 mm. The light then passes through a PBS, QWP, and HWP
to generate a well-defined uniform input polarization state. We optionally
insert spatially varying birefringent elements here in order to create
a non-uniform polarization state.

The light passes through the first 50:50 NPBS and is then incident
on the right half of the SLM (Hamamatsu X10468-07, $792\times600$
pixels, pixel size $20\times20$ $\mu$m, 256 phase increments) normal
to its surface. Here, the first polarization rotation, $\hat{R}(\zeta_{1,ij},\mathbf{S_{1}})$,
occurs. The light then reflects from this half of the SLM and the
first NPBS and is directed by silver mirrors through a HWP with its
optical axis oriented at $22.5^{\circ}$ from the horizontal. The
light transmits through a second 50:50 NPBS before its incidence on
the left half of the SLM, where a second polarization rotation occurs.
When this rotation is considered with the HWP, this rotation is $\hat{R}(\zeta_{2,ij},\mathbf{S_{2}})$.
Finally, the light reflects from the SLM and the second NPBS and heads
towards the Stokes polarimetry apparatus.

The two-step scheme is capable of tailoring highly non-uniform output
polarization profiles. To do so, the SLM must impart a large phase
gradient across the light field, which, in turn, creates components
of the field travelling at large angles. In order to retain this angular
spread, a $4f$ imaging system is used to image the field on each
half of the SLM. Each 4f imaging system consists of a pair of lenses
($f=100$ mm, diameter $2.5$ cm, planoconvex) separated by $2f$.
The first and last lens are positioned a distance $f$ away from the
object and image plane, respectively. The first 4f system ($l_{4},$
$l_{5}$) images the field at the right half of the SLM onto the left
half of the SLM. The second $4f$ system ($l_{6},$ $l_{7}$) images
the field at the left half of the SLM onto an image sensor. In the
appendix, we detail how we compensate for image flips caused by the
mirrors and lenses.

Following this last lens, we characterize the output polarization
by conducting standard Stokes polarimetry using a QWP, HWP, and PBS
\cite{goldstein:03}. Combined with a digital image sensor (Basler
aca1600-20gm, $1626\times1236$ pixels, pixel size of $4.4\times4.4$
$\mu$m, 12-bit bit depth, zero gain, 0.65 s exposure), this allows
us to measure each Stokes component at each sensor pixel, thereby
determining the polarization state point-by-point across the light
field. In order to reduce the effect of waveplate retardance errors,
we eliminate the second HWP that is nominally required for a rotation
about $\mathbf{S_{2}}$. Its absence can be compensated for in the
data analysis by simply swapping the definitions of the Stokes components
$s_{1}$ and $s_{2}$ with one another. All waveplates are true zero-order
with a design wavelength of 633 nm.

\section{Experimental Setup Calibration}

\label{sec:calibration}

\subsection{SLM Phase-Grey Calibration}

\begin{figure}[h!]
\centering \includegraphics[width=1\textwidth]{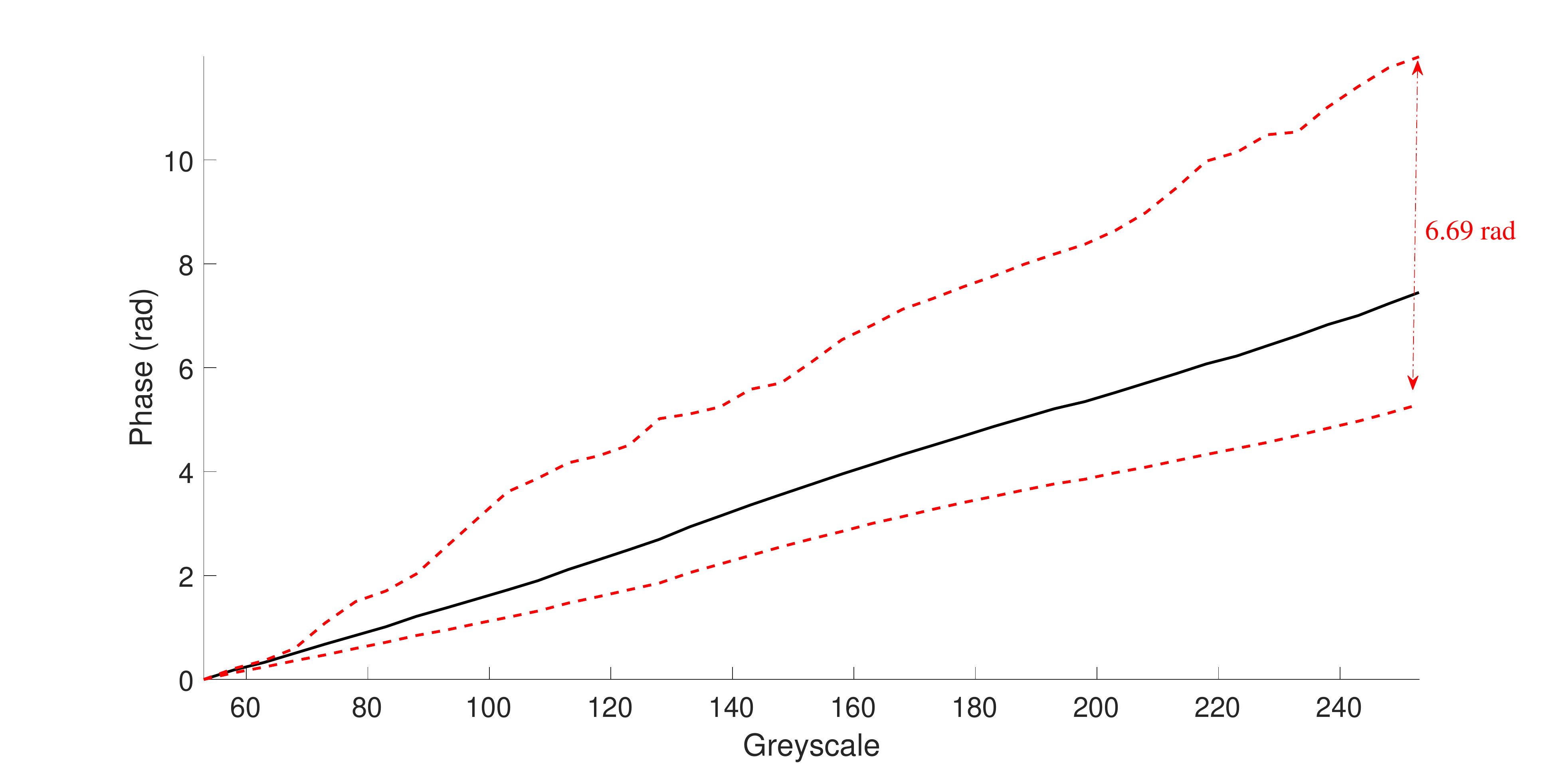}
\captionsetup{width=\textwidth} \caption{The phase-grey relationship is plotted for pixels within the $1/e^{2}$
beam waist across the wave front. The red dashed curves correspond
to the pixels which had the largest and smallest changes in phase
for a change in greyscale, and the black curve denotes the average
phase-grey relationship amongst all beam waist pixels. A maximum phase
discrepancy of 6.69 rad (roughly $120$ greyscale levels), occurs
at a maximum greyscale value of 253.}
\label{fig:phasegray} 
\end{figure}

The amount of phase that the SLM imparts at a given pixel is directly
proportional to an 8-bit value at that pixel. Since the SLM is electronically
controlled by a standard digital video signal, we refer to this as
a greyscale. The phase to greyscale relationship is nominally given
by the manufacturer as a linear function, $\zeta_{ij}\left(g_{ij}\right)=Cg_{ij}$,
where $g_{ij}\in[0,255]$ is the greyscale value at pixel $(i,j)$
and $C=2\pi/118$ is the nominal phase-grey proportionality constant
provided by the manufacturer. However, we have found that the value
of $C$ varies from one pixel to the next, as seen in Fig. \ref{fig:phasegray}.
To accurately program the SLM, an independent calibration of each
illuminated pixel was carried out. This was accomplished by measuring
the rotation in polarization at discrete points $(i,j)$ of the output
light field as a function of greyscale on the corresponding pixels
$(i',j')$ on the SLM. We used look up tables (LUTs) to define this
exact phase-grey relationship $\zeta_{ij}\left(g_{ij}\right)$ for
each SLM pixel. 

Since there are two incidences on the SLM, two separate calibration
runs were carried out. Performing the calibration of either side independently
is not necessary, but is useful in what it reveals about the setup,
as explained in the appendix. %

\subsection{Phase Offset Determination}

\label{subSec:phaseCal} It is possible that at zero greyscale, the
SLM still imparts a phase-shift between $\ket{H}$ and $\ket{V}$.
Moreover, every mirror and NPBS can impart a phase-shift between $\ket{H}$
and $\ket{V}$, due to the oblique (typically 45$^{\circ}$) beam
incidence angle. Since these and the intrinsic SLM phase are both
rotations around $\mathbf{S_{1}}$, their effect can simply be summed
and then compensated by a greyscale pattern on the SLM. However, between
the two SLM reflections is a HWP, which induces a rotation about $\mathbf{k}=(\mathbf{S_{1}}+\mathbf{\mathbf{S}_{2}})/\sqrt{2}$.
Thus, we determine a phase offset for each half of the SLM (before
and after the HWP). We compensate for this phase offset pixel-by-pixel
by applying a greyscale offset-compensation matrix, $A_{ij}$, on
each half of the SLM.

\begin{figure}[h!]
\centering \includegraphics[width=0.6\textwidth]{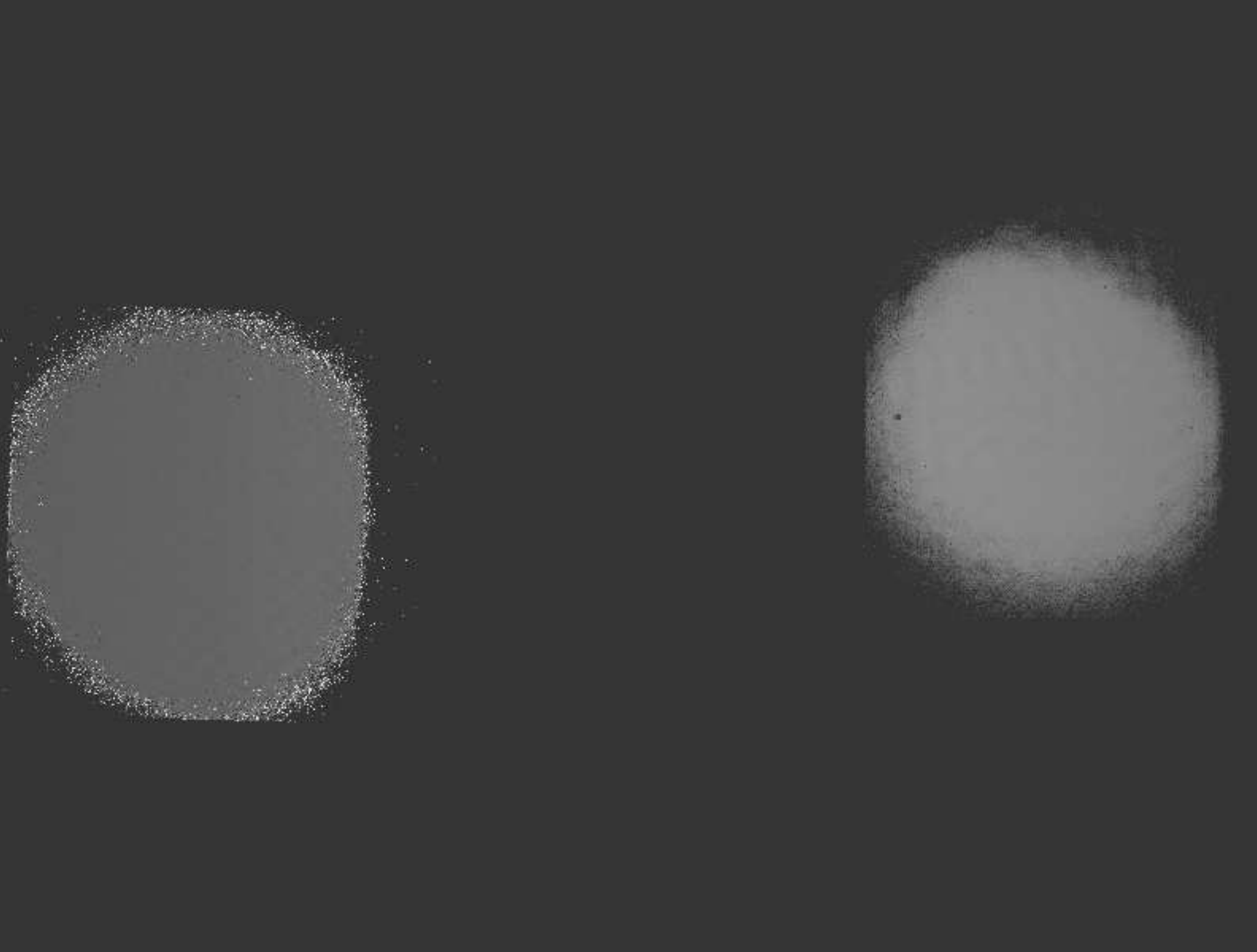} \captionsetup{width=0.8\textwidth}
\caption{The greyscale offset-compensation matrix, $A_{ij}$ for each half
of the SLM. Black pixels (zero greyscale) represent zero imparted
phase and white pixels (255 greyscale) represent a maximum imparted
phase. On average, the compensation requires roughly 2.25 rad of phase
about $\mathbf{S_{2}}$ and 0.86 rad of phase about $\mathbf{S_{1}}$.}
\label{fig:graycal} 
\end{figure}

This offset-compensation is found as follows: a uniform, diagonally
polarized state, $\mathbf{\mathbf{S}}_{ij}=[0,1,0]$, is used as the
input state to the setup and the output polarization $\mathbf{S}'_{ij}$
is measured at each pixel. The change in polarization from $\mathbf{S}{}_{ij}$
to $\mathbf{S}'_{ij}$ is described by an angular displacement on
the Poincaré sphere. This angle is decomposed into two components;
the first is a rotation of angle $\phi_{1}$ about $\mathbf{S_{1}}$.
The second component is a rotation of angle $\phi_{2}$ about $\mathbf{S_{2}}$.
The phases $\phi_{1}$ and $\phi_{2}$ are given by Eq. (\ref{eq:zeta1})
and Eq. (\ref{eq:zeta2}) in the appendix, respectively. The compensating
phase is merely $\zeta_{1,ij}=2\pi-\phi_{1}$ and $\zeta_{2,ij}=2\pi-\phi_{2}$
for the first and second halves of the SLM, respectively. These phases
vary from one pixel to the next. The corresponding greyscales for
all the pixels comprise matrix $A_{ij}$. The matrices for the two
SLM halves, positioned properly, constitute the image that is shown
in Fig. \ref{fig:graycal}. This image is added in greyscale (modulo
255) to all greyscale images used to implement an arbitrary two-step
transformation.

With this compensation in place, the system should not modify the
input polarization. Thus, the output polarization $\bar{\mathbf{S}}'$
should be identical to any chosen input polarization. We test this
with the nominal uniform input polarization state, $\mathbf{S}_{\mathrm{ideal}}=[0,1,0]$.
This is our target state. To distinguish the performance of the polarization
system from our ability to produce and measure polarizations, we also
characterize the nominal input state by sending it directly to the
polarimetry setup. We call the latter the experimental input state.
In Table \ref{table:phaseCalCheck}, we compare the target state to
the experimental input, the output without compensation, and the output
with compensation. When the phase compensation is used, the fidelity
of $\bar{\mathbf{S}}'$ with respect to the target input state improves
by a percent difference of 85\%, and the uniformity improves by a
percent difference of 5\%. Both measures indicate that the compensation
significantly improves the system performance.

\begin{table}[h!]
\centering \caption{Performance evaluation of phase-offset compensation.}
{\global\long\def\arraystretch{1.2}
\begin{tabular}{|c|c|c|c|}
\hline 
\textbf{State}  & \textbf{Avg. Stokes Vector,} $\bar{\mathbf{S}}_{\mathrm{exp}}$  & \textbf{Fidelity, $F$}  & \textbf{Uniformity, }$U$\tabularnewline
\hline 
input, target  & $\mathbf{\mathbf{S}_{\mathrm{ideal}}}$ = {[}0,1,0{]}  & 1  & 1\tabularnewline
\hline 
\multicolumn{1}{|l|}{experimental input} & $\bar{\mathbf{S}}$ = {[}0.075, 0.986, 0.012{]}  & 0.993  & 0.988\tabularnewline
\hline 
\multicolumn{1}{|l|}{w/o compensation} & $\bar{\mathbf{S}}'$ = {[}0.244, 0.056, -0.889{]}  & 0.528  & 0.924\tabularnewline
\hline 
\multicolumn{1}{|l|}{with compensation} & $\bar{\mathbf{S}}'$ = {[}-0.114, 0.957, 0.071{]}  & 0.978  & 0.966\tabularnewline
\hline 
\end{tabular}}\label{table:phaseCalCheck} {\footnotesize{}{}{}}{\footnotesize \par}

{\footnotesize{}{}{*} The fidelity is $F=F(\bar{\mathbf{S}}{}_{\mathrm{exp}},\mathbf{S}{}_{\mathrm{ideal}})$.} 
\end{table}

\begin{figure}[h!]
\centering \includegraphics[width=1\textwidth]{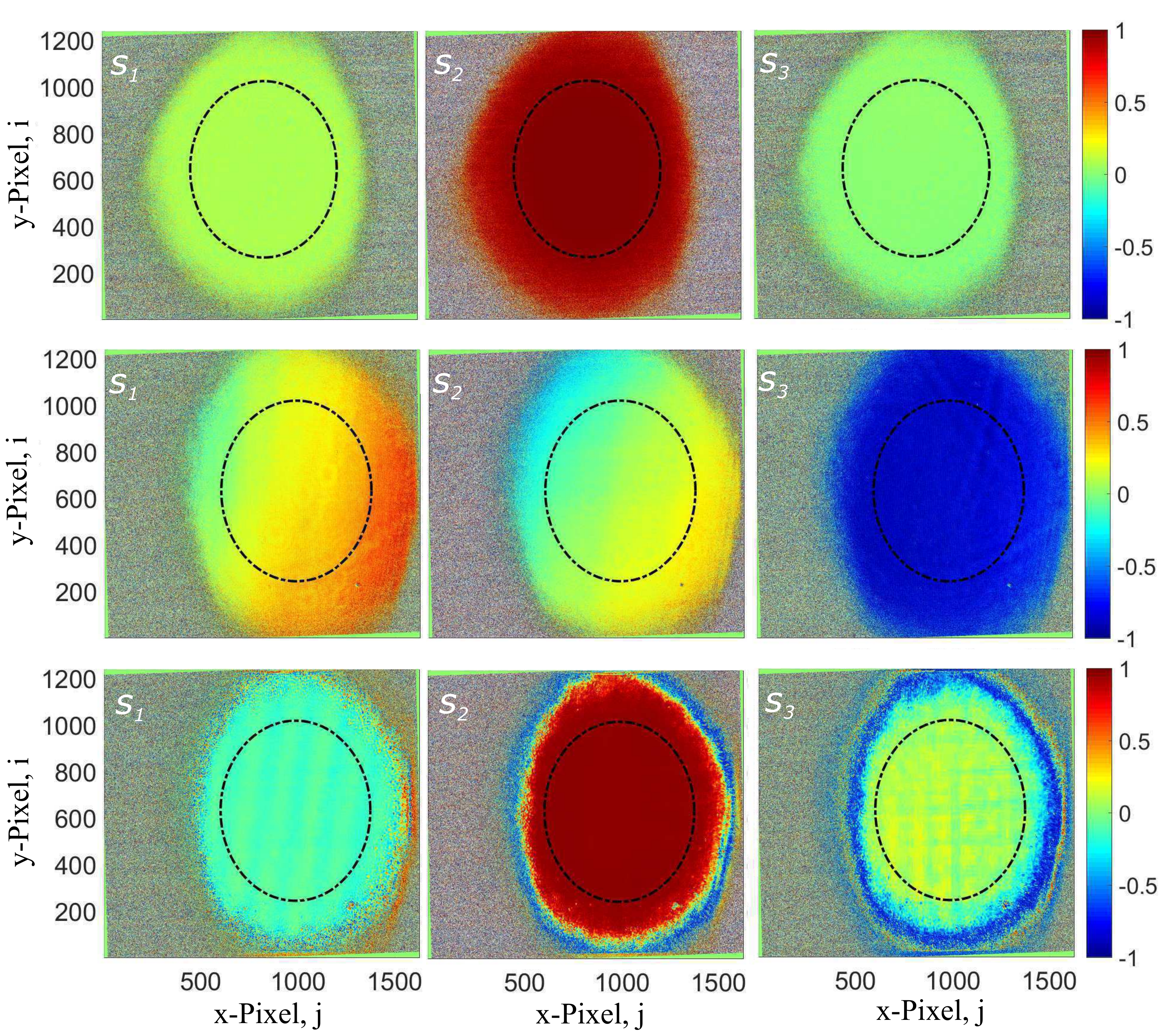}
\captionsetup{width=1\textwidth} \caption{Performance of the phase offset compensation. Each density plot shows
the Stokes components across the wave front for $s_{1}$, $s_{2}$,
and $s_{3}$ organized from left to right, and the black dotted line
corresponds to the FWTM beam waist. The first row contains density
plots for an experimentally measured input state $\mathbf{S}_{ij}$.
Ideally, $\mathbf{S}_{ij}=\mathbf{S}_{\mathrm{ideal}}=[0,1,0]$. This
data characterizes the performance of our polarimetry and input polarization
state production. The other rows contain density plots for the measured
output state for the input $\mathbf{S}_{ij}$ when there is no phase
compensation (second row) and when there is phase compensation (third row).
If the phase compensation was completely perfect, the first and third
rows would be identical. }
\label{fig:CompensationComparison} 
\end{figure}

\section{Experimental Results}

\label{sec:results}

\subsection{Uniform to Non-Uniform Transformations}

\label{sec:uni2nonuni} To demonstrate the capability of performing
non-uniform polarization transformations, in this section we transform
a uniformly $\ket{D}$-polarized beam to a different output polarization
in each quadrant of the light field. We choose target states $\mathbf{S}'_{\mathrm{ideal}}$
that constitute a set forming a symmetric, informationally complete
positive-operator valued measure (SIC-POVM) \cite{caves:04}. These
are known to be optimal for performing quantum polarization state
tomography \cite{caves:04}. When the SIC-POVM states are plotted
on the Poincaré sphere, the Stokes vector tips point to the vertices
of a tetrahedron. Since they are equally distributed around the Poincaré
sphere, our ability to produce them is a good indication that any
arbitrary output polarization can be created.

\begin{table}[h]
\setlength{\tabcolsep}{5pt} \caption{Performance of uniform to non-uniform transformations. }
{\global\long\def\arraystretch{1.2}
\begin{tabular}{|c|c|c|c|c|}
\hline 
 & \textbf{Quadrant 1}  & \textbf{Quadrant 2}  & \textbf{Quadrant 3}  & \textbf{Quadrant 4} \tabularnewline
\hline 
$\mathbf{S}'_{\mathrm{ideal}}$  & $\frac{-1}{\sqrt{3}}[1,1,1]$  & $\frac{1}{\sqrt{3}}[-1,1,1]$  & $\frac{1}{\sqrt{3}}[1,1,-1]$  & $\frac{1}{\sqrt{3}}[1,-1,1]$\tabularnewline
\hline 
$\bar{\mathbf{S}}'_{\mathrm{exp}}$  & $[-0.44,-0.70,-0.51]$  & $[-0.50,0.58,0.51]$  & $[0.53,0.41,-0.48]$  & $[0.49,-0.43,0.61]$\tabularnewline
\hline 
$U$  & 0.93  & 0.88  & 0.88  & 0.95\tabularnewline
\hline 
\textbf{$F$}  & 0.96  & 0.94  & 0.93  & 0.97 \tabularnewline
\hline 
\end{tabular}} \label{table:uni2nounires} {\footnotesize{}{}{}{*}The uniformity
is $U$. The fidelity is $F=F(\bar{\mathbf{S}}'_{\mathrm{exp}},\mathbf{S}'_{\mathrm{ideal}})$,
where $\bar{\mathbf{S}}'_{\mathrm{exp}}$ is the measured state and
$\mathbf{S}'_{\mathrm{ideal}}$ is the target state. Note that $\frac{1}{\sqrt{3}}=0.58.$} 
\end{table}

\begin{figure}[h!]
\centering \includegraphics[width=1\textwidth]{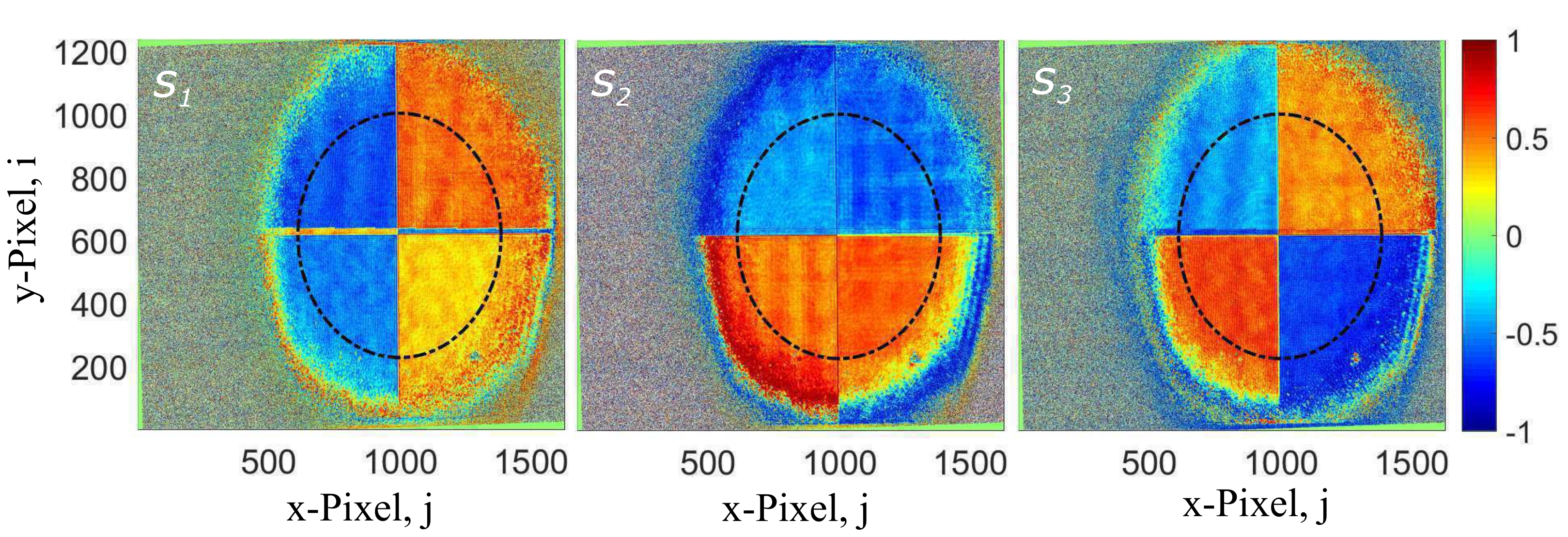}
\caption{{\footnotesize{}{}The Stokes components for an ouput light field
that contains polarization states equally distributed around the Poincaré
sphere. Each quadrant is transformed to contain one such SIC-POVM
state. For the purpose of Table \ref{table:nonuni2uni}, the quadrants
are numbered starting from the bottom left and moving clockwise.}}
\label{fig:tetrabeam} 
\end{figure}

\begin{figure}[h!]
\centering \includegraphics[width=0.5\textwidth]{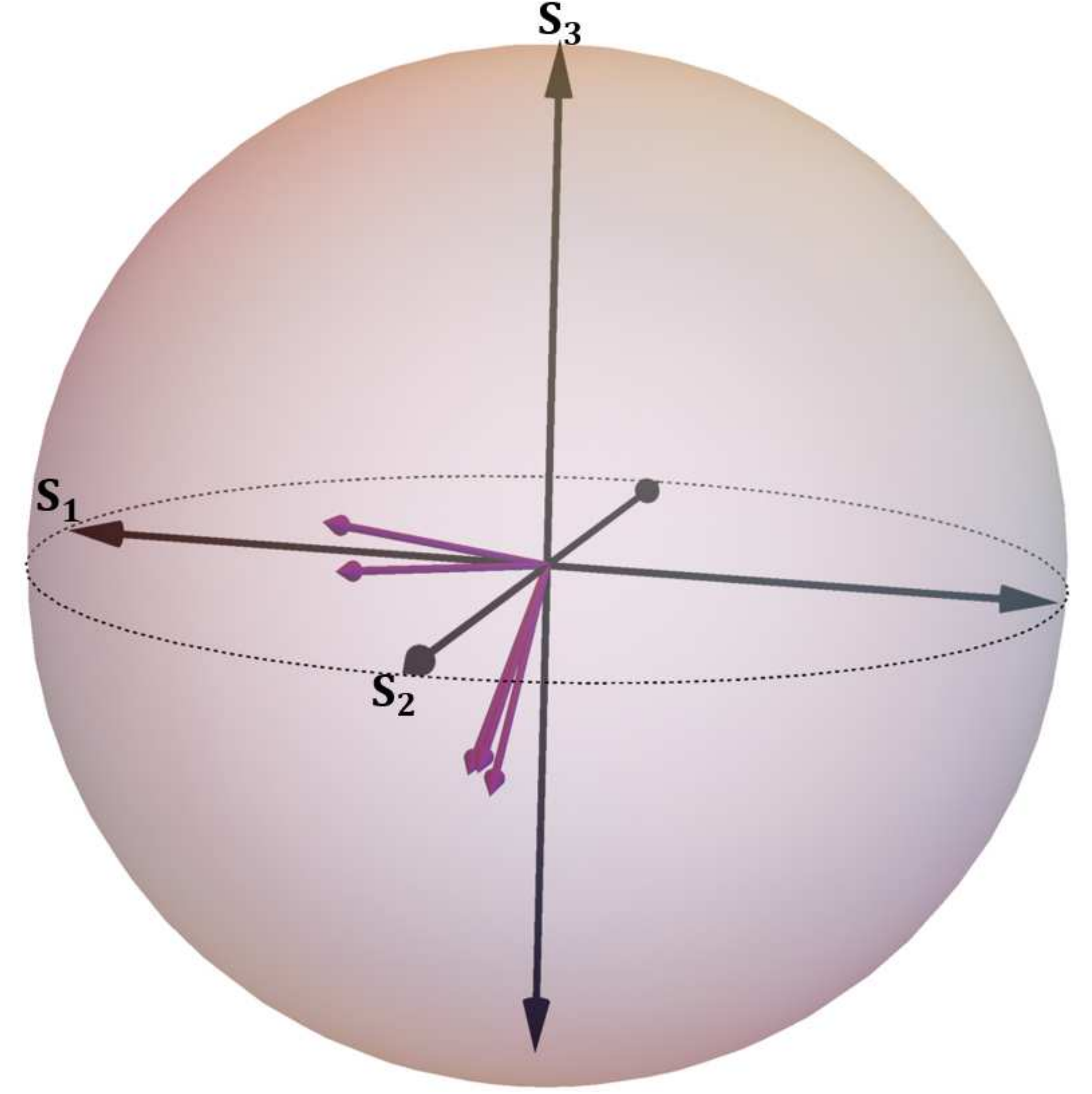}
\captionsetup{width=0.6\textwidth} \caption{The axes of rotation are plotted for five distinct SLM pixels in the
FWTM beam waist. These particular axes exhibit the greatest deviations
from the ideal axis of rotation, $\mathbf{S_{2}}$, and this deviation
can be as large as 0.29 rad.}
\label{fig:pixax} 
\end{figure}

The results of this transformation are given in Table \ref{table:uni2nounires}.
The experimental average Stokes vector for each output quadrant $\bar{\mathbf{S}}'_{\mathrm{exp}}$
are calculated over a circular sector (i.e., wedge) $A$ wholly within
that quadrant. The fidelities of the four quadrants with respect to
their target polarization states range from 0.93 to 0.97 and the uniformities
range from 0.88 to 0.95. While all the states are produced reasonably
well, the uniformity is significantly lower in some quadrants than
others. Our investigations into this effect revealed that each pixel
actually has a different axis of rotation in the Poincaré sphere,
as seen in Fig. \ref{fig:pixax}. In some cases, this difference is
quite large and can change the intended transformation of one pixel
by a phase corresponding to as much as ten greyscale increments. This
error will also be more profound for some states than others. In particular,
the axes differ in such a way that any state with a negative $s_{2}$
component will be constructed more accurately than states with a positive
$s_{2}$ component.

Another cause for imperfect results is the presence of a grid-like
pattern in polarization that shows up in the density plots of the
Stokes parameters (e.g., Fig. \ref{fig:CompensationComparison}).
This grid-like polarization effect is thought to be a diffraction
effect from the pixels of the SLM. One can see in Fig. \ref{fig:graycal}
that this pattern is noticeable at the greyscale level.

\subsection{Painting with Polarization}

While this non-uniform transformation successfully resulted in the
desired state, only four distinct transformations were performed at
once. This is a small number of transformations when one considers
that each pixel of the SLM can transform light independently. To demonstrate
this versatility we use the setup to implement \emph{polarization
imaging}. Here, full images can be encoded in polarization that are
hidden to humans, a species whose vision is mostly polarization insensitive,
but can be revealed through Stokes polarimetry and graphical representations
of polarization.

\begin{figure}[h!]
\centering \includegraphics[width=1\textwidth]{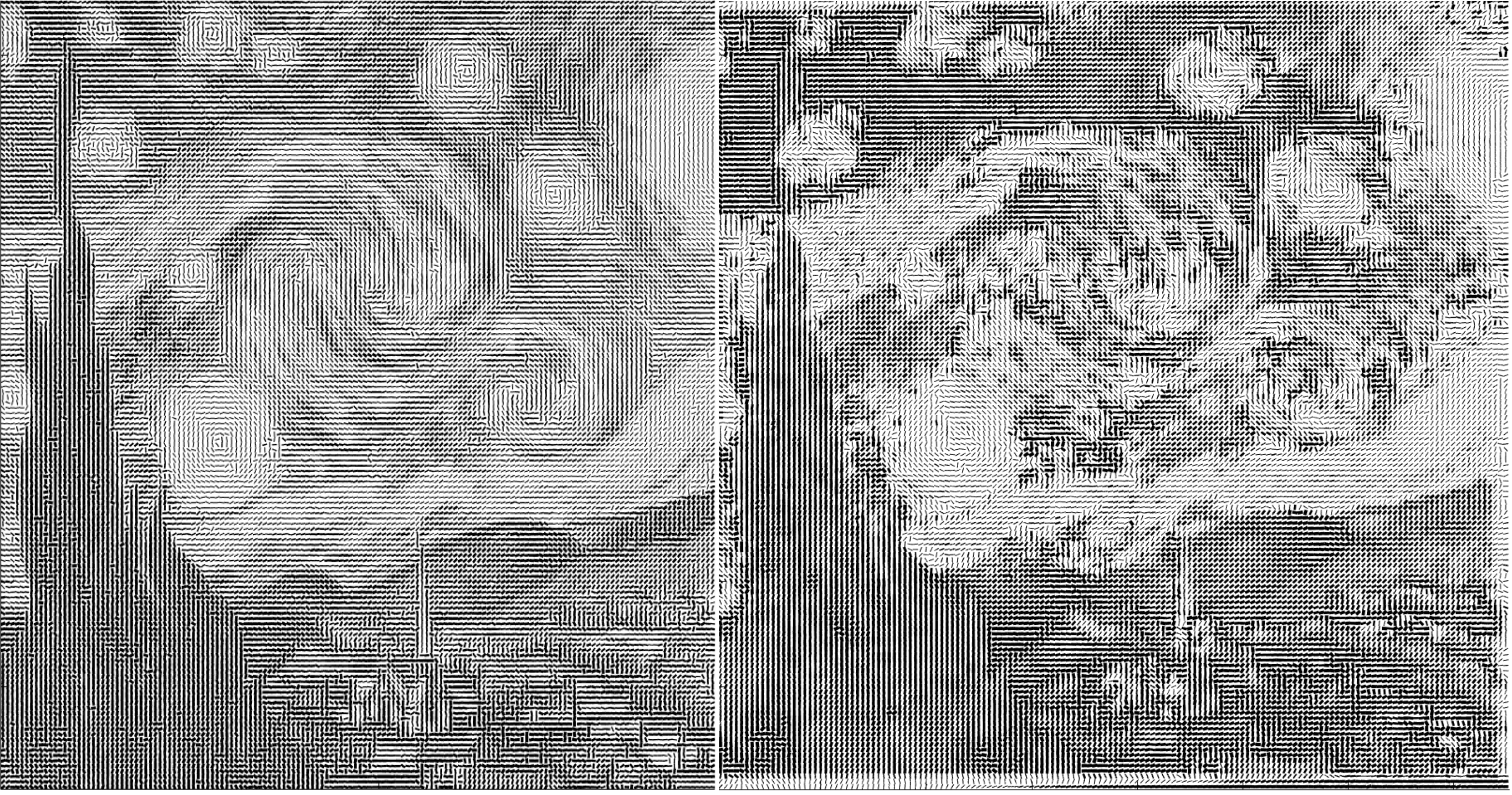}
\caption{Van Gogh's painting \emph{Starry Night}, painted in polarization.
On the left is the simulated target image, on the right is the experimental
result. In each region, the line segment angle corresponds to the
dominant polarization direction. The thickness of the line segment
is set by the ellipticity of the polarization in that region of the
light field.}
\label{fig:starryNight} 
\end{figure}

Here, a cropped image of Van Gogh's painting \emph{Starry Night} was
encoded in polarization. This painting is ideal for this because of
its characteristic feature, the distinctly clear brush strokes that
form swirls reminiscent of a vector field. Each region $(i,j)$ of
the painting is converted to a target polarization state $\mathbf{S}'_{ij}$.
The darkness $D(i,j)$ of the region sets the polarization ellipticity
through $s_{3,ij}=1-2(D(i,j)/D_{max})$, where $D_{max}$ is the maximum
darkness in the painting. The angle $\psi(i,j)$ of the brush stroke
in the region is converted to the dominant polarization direction
through $s_{1,ij}=\sqrt{1-s_{3,ij}^{2}}\cos(2\psi(i,j))$ and $s_{2,ij}=\sqrt{1-s_{3,ij}^{2}}\sin(2\psi(i,j))$.
With this target $\mathbf{S}'_{ij}$ and a known input polarization
$\mathbf{S}{}_{ij}$, the two-step scheme can be used to imprint a
polarization image of \emph{Starry Night} on a light field. We do
so with a nominal uniformly diagonally polarized input field, $\mathbf{S}=[0,1,0]$.
A species that could see linear polarization (e.g., octopus or mantas
shrimp) might be able to perceive this image directly. Conceivably,
they would see something similar to the left image in Fig. \ref{fig:starryNight},
where we have converted the polarizations $\mathbf{S}'_{ij}$ back
to line segments at angles $\psi(i,j)$. The darkness $D(i,j)$ sets
the thickness of the line segment. We, however, must use our Stokes
polarimetry setup to perceive the image. We similarly convert the
experimentally measured Stokes vectors to the right image in Fig.
\ref{fig:starryNight}. It is, unmistakably, \emph{Starry Night,}
therein demonstrating our ability to effectively paint with polarization.

\subsection{Non-Uniform to Uniform Transformations}

\label{sec:nonuni2uni} 
\begin{figure}[h!]
\centering \includegraphics[width=1\textwidth]{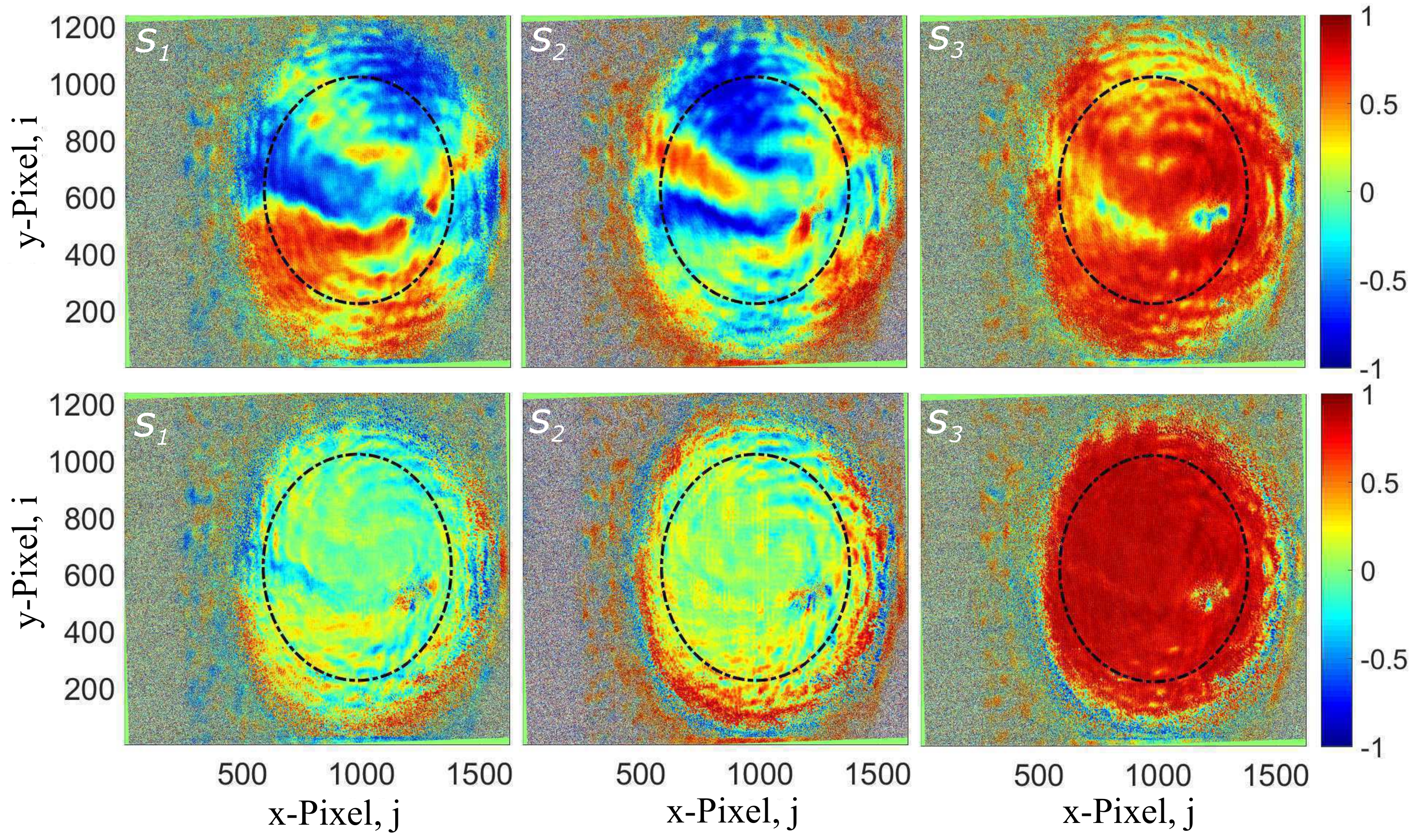}
\caption{{\footnotesize{}{}Correcting polarization aberrations. We transform
a light field with a highly non-uniform polarization $\mathbf{S}_{ij}$ (top row)
to a more uniform polarization $\mathbf{S}'_{ij}$ (bottom row). }}
\label{fig:qplateBase} 
\end{figure}

Performing non-uniform-to-uniform transformations
may also offer applications in \emph{beam healing}, where polarization
aberrations and undesired non-uniformities can be corrected for by
bringing a light field back to a uniform polarization state. To demonstrate
that the setup is capable of these transformations, a liquid crystal
cell with a spatially varying optic axis orientation was added between
the input polarization state preparation stage and the two-step scheme
setup. This transforms a uniform, circularly polarized state to a
highly non-uniform polarization input state. With only the phase-offset
compensation in place, we measured this non-uniform polarization profile.
We took this pre-corrected state as our input $\mathbf{S}{}_{ij}$.
We then used the setup to transform to the uniformly polarized target
state, $\mathbf{S}'_{\mathrm{ideal}}=[0,0,1]$. The experimentally
measured pre-corrected state $\mathbf{S}{}_{ij}$ and corrected state
$\mathbf{S}'_{ij}$ are respectively shown in the top and bottom of
Fig. \ref{fig:qplateBase}. To the eye, the uniformity of the light
field is considerably improved. As summarized in Table \ref{table:nonuni2uni},
this impression is confirmed by the nearly 40\% increase in uniformity
of the light field. This demonstrates that the two-step scheme can
be used to heal polarization aberrations in beams.

\begin{table}
\centering \caption{Homogenizing a highly non-uniformly polarized light field towards
a target of $\mathbf{S}'_{\mathrm{ideal}}=[0,0,1]$.}
\begin{tabular}{|l|p{2cm}|p{2cm}|p{2cm}|}
\hline 
\textbf{Polarization field}  & \textbf{Avg. Stokes vector,} $\bar{\mathbf{S}}{}_{\mathrm{exp}}$  & \textbf{Fidelity, $F(\bar{\mathbf{S}}_{\mathrm{exp}},\mathbf{S}'_{\mathrm{ideal}})$}  & \textbf{Uniformity, }$U$\tabularnewline
\hline 
Pre-corrected state, $\mathbf{S}{}_{ij}$  & $\bar{\mathbf{S}}$  & 0.799  & 0.632 \tabularnewline
\hline 
Corrected state, $\mathbf{S}'{}_{ij}$  & $\bar{\mathbf{S}}'$  & 0.933  & 0.869 \tabularnewline
\hline 
\textbf{\% Difference}  &  & \textbf{17 \%}  & \textbf{38 \%} \tabularnewline
\hline 
\end{tabular}\label{table:nonuni2uni} 
\end{table}

\section{Conclusion}

We have shown that two-step polarization transformations can be used
to produce arbitrary, spatially varying polarization states from a
known polarization state. In particular, polarizations in each quadrant
of a wave front were produced as desired with fidelities upwards of
0.93. It was also shown that these transformations can be used to
paint with polarization, where Van Gogh's painting \emph{Starry Night}
was embedded in the polarization of a light field through the use
of elliptical polarization states. Finally, it was shown that these
transformations can be used to `heal' the polarization of a light
field, where a light field with an initial uniformity of 0.632 was
transformed to have a uniformity of 0.869.

Future directions of this work include using the system to perform
transformations that bring a non-uniform state to another non-uniform
state. Such transformations would be useful in spatially multiplexed
polarization-based quantum cryptography. Another noteworthy application
is the non-diffractive patterning of passive liquid crystal devices.
Standard techniques used to pattern liquid crystal devices use photoalignment
methods involving a digital micro-mirror device (DMD) for micro-lithography
\cite{chigrinov:12}. In them, arbitrarily complex liquid crystal
patterns are created by sequentially exposing various regions of a
dye or polymer to various polarizations. The two-step scheme would
require only a single exposure, thereby increasing production throughput.
These patterned liquid crystal devices (e.g., q-plates) are static.
They are each designed to generate select subsets of vector beams.
In contrast, the two-step scheme can create any arbitrary spatial
polarization, which itself can be changed at a rate of 60 Hz. If transmissive
SLMs are improved, one may reduce the complexity of the setup by using
transmissive devices in place of reflective ones in order to avoid
stray reflections and their associated spurious optical interference.
As well, transmissive SLMs would make experiments based on three successive
phase modulations more feasible. These could implement even more general
transformations than those presented here, transformations with an
arbitrary Poincaré sphere axis and angle of rotation at each pixel
\cite{sit:17}.

\section*{Funding}

This research was undertaken, in part, thanks to funding from the
following organizations: Canada Research Chairs, NSERC Discovery,
Canada Foundation for Innovation, Canada First Research Excellence
Fund, and Canada Excellence Research Chairs program.

\appendix

\section*{Appendix}

\global\long\def\thesubsection{A\arabic{subsection}}

\subsection{Alignment and Mapping of the SLM to the CCD}

\label{subSec:ImProc} 
\begin{figure}[h!]
\centering \includegraphics[width=1\textwidth]{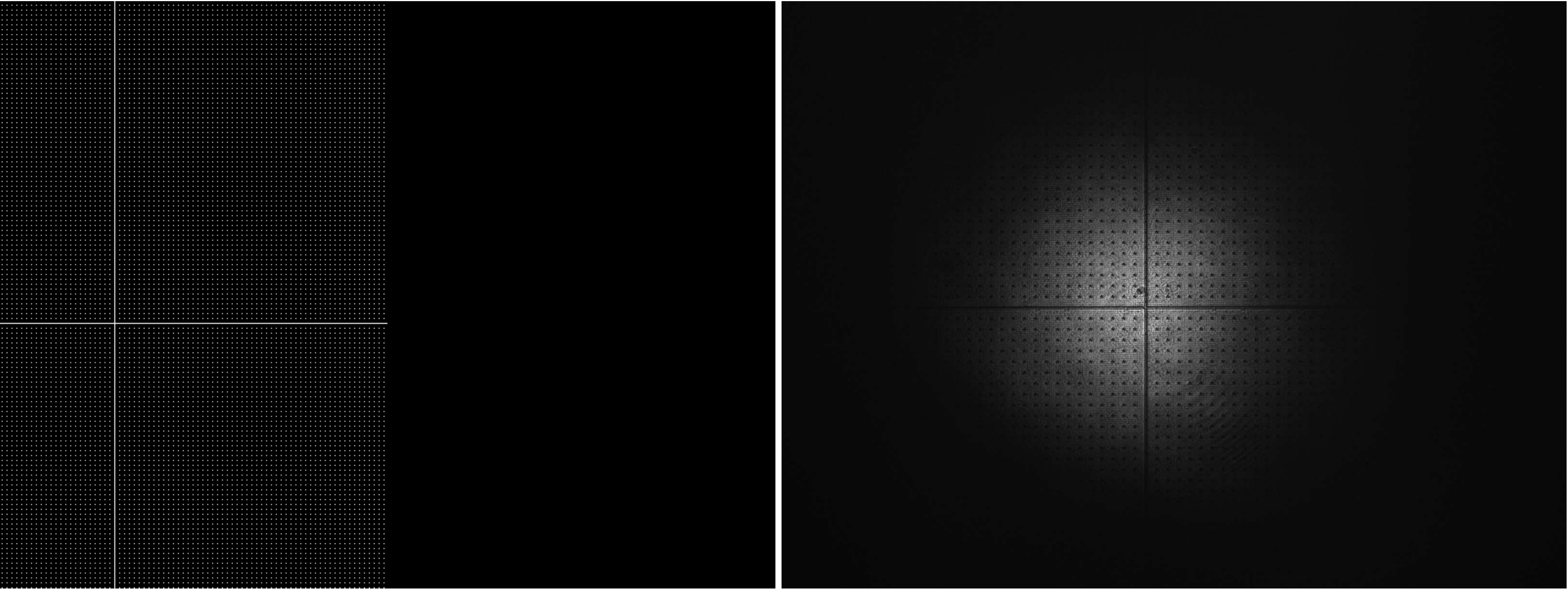} \caption{The fiducial greyscale pattern on the left half of the SLM (left)
and the resulting image on the CCD (right).}
\label{fig:fiducial} 
\end{figure}

For the two-step scheme to function we must
program particular SLM pixels to modify particular regions of the
light field. We now describe an alignment procedure and mapping between
the pixels illuminated by the light field on each of the two separate
incidences of the SLM, $(i_{1}^{SLM},j_{1}^{SLM})$ and $(i_{2}^{SLM},j_{2}^{SLM})$,
and the pixels of the CCD camera, $(i^{CCD},j^{CCD})$, which are
at position $(x_{j},y_{i})$ of the wave front. The alignment and
mapping is accomplished by displaying a fiducial greyscale pattern
on each half of the SLM. In the left of Fig. \ref{fig:fiducial},
we show this pattern in the case when it is present only on the left
half of the SLM. After passing through the SLM part of the setup,
the light travels through the Stokes polarimetry apparatus to form
a clear image of the pattern on the CCD, as shown on the right in
Fig. \ref{fig:fiducial}. Our mapping function, this vector formula,
compensates for four issues: scaling, offset, reflection, and rotation:
\begin{equation}
\begin{bmatrix}i_{k}^{SLM}-i_{k,0}^{SLM}\\
j_{k}^{SLM}-j_{k,0}^{SLM}
\end{bmatrix}=\begin{bmatrix}\cos\theta_{k} & -\sin\theta_{k}\\
\sin\theta_{k} & \cos\theta_{k}
\end{bmatrix}\begin{bmatrix}\lfloor m_{k,y}\left(i^{CCD}-i_{0}^{CCD}\right)\rfloor\\
\lfloor m_{k,x}\left(j^{CCD}-j_{0}^{CCD}\right)\rfloor
\end{bmatrix}.\label{eq:mapping}
\end{equation}
In the following sections, we explain this mapping, define the parameters
in it, and describe how they were determined.

\subsubsection{Image Offset}

The pattern on each half of the SLM has an offset in reference to
the CCD array. The image of the bright central cross-hair in Fig.
\ref{fig:fiducial} from the left and right halves of the SLM were
manually aligned to the center of the light field by shifting the
corresponding greyscale patterns. The exact position of the cross-hair
on the CCD is used to set an origin for the CCD $(i_{0}^{CCD},j_{0}^{CCD})$,
as well as the two SLM offsets $(i_{k,0}^{SLM},j_{k,0}^{SLM}$), where
$k=1,2$ corresponding to the left and right half of the SLM. This
fixes all the pixel offsets in Eq. (\ref{eq:mapping}).

\subsubsection{Image Scaling}

While the $4f$ imaging nominally has a magnification of one, it might
deviate from this experimentally. As well, the pixel sizes of the
CCD and SLM are different. We account for these issues with two components
in our mapping, scaling and resampling. The scaling is accounted for
by the $m_{k,y}$ and $m_{k,x}$ magnification factors in Eq. (\ref{eq:mapping}),
for the first and second SLM halves, $k=1,2$. We determine these
factors using the grid of dots in the fiducial pattern. To accurately
determine the scaling we take a Fourier transform of the CCD image
to determine the spatial frequency of the pattern. With this, we determine
that $m_{y}=0.209$ and $m_{x}=0.208$. Identical results were obtained
for both halves of the SLM, so we have dropped the $k$ subscript
for now. According to the pixel size specifications of the CCD and
SLM, and if we had an optical magnification of one, we would expect
$m_{x}=m_{y}=0.22$. This is $5.5\%$ larger than the experimental
values.

Since the magnification factors are less than one, there are $1/m_{x}m_{y}$
CCD pixels for each SLM pixel. This many-to-one mapping, i.e., resampling,
is accomplished using a floor function, denoted by the $\lfloor\rfloor$
symbols in Eq. (\ref{eq:mapping}). The floor function rounds to the
nearest lower integer. In doing so it fixes only one CCD pixel to
correspond to each SLM pixel. An alternative would be to average over
a region of $1/m_{x}m_{y}$ CCD pixels to derive one effective CCD
pixel for each SLM pixel. Standard imaging specific resampling functions
(e.g., bicubic) could also be used, though these are less transparent
in their functioning. All three methods were attempted and flooring
was chosen as the most robust in its functioning.

\subsubsection{Image Reflections}

There are multiple sources of image flips (e.g., invert horizontally,
$x\rightarrow-x$) in the system. These flips can occur between incidences
on the SLM and between the SLM and CCD. Every reflection from a mirror,
beamsplitter or the SLM will invert $x$. Each $4f$ lens system inverts
both $x$ and $y$. Lastly, the orientation (i.e., the default direction
of increasing pixel number) of the CCD sensor can be different from
that of the SLM. Mathematically, this translates to negation of either
or both magnification factors, $m_{k,y}$ and $m_{k,x}$ for Eq. (\ref{eq:mapping})
for the first and second SLM halves, $k=1,2$. These negations can
be determined by placing non-symmetric images (e.g., the letter F)
on each of the two SLM halves and simply observing on the CCD in which
directions they have each been flipped. In our system, only the second
half of the SLM must be flipped. Consequently, the magnification factor
$m_{2,x}=-0.208$ has been negated.

\subsubsection{Image Rotation}

The image can rotate if the light propagation axis is not parallel
with the optical table. This deviation from parallel occurs in some
places in our setup, as is evident by the offset between the left
and right greyscale patterns in Fig. \ref{fig:graycal}. We incorporated
this rotation in Eq. (\ref{eq:mapping}) with a standard rotation
matrix that rotates the $x-y$ plane by an angle of $\theta_{k}$.
This rotation can be determined directly by the orientation of the
cross-hair or through a Fourier transform of the grid of dots in the
fiducial pattern. From this, we find that there is no rotation between
the second half of the SLM and the CCD. However, the fiducial on the
first half on the SLM appears rotated relative the CCD sensor edges
by $0.0321$ radians. We set $\theta_{1}=0$ and $\theta_{2}=0.0321$
radians so that light incident on a pixel on the second half of the
SLM is reflected from the corresponding pixel on the first have of
the SLM. i.e., $i_{1}^{SLM}=i_{2}^{SLM}$ and $j_{1}^{SLM}=j_{2}^{SLM}$.
The result is that fiducials that are aligned with the pixel columns
on the first and second half of the SLM will appear on CCD exactly
aligned (i.e., superposed). However, they will be rotated with respect
to CCD sensor axes. We rotate the final image is back by $-\theta_{2}$
during post-processing to ensure the polarization pattern (and fiducial)
appears upright as intended. In any of the experimental density plots
of the Stokes components, one can see signs of this rotation at the
plot edges.

\subsubsection{Nonlinear Image Distortion}

We also used the fiducial grid to look for evidence of a nonlinear
component to our mapping, such as pincushion or barrel distortion.
Such distortions are not included in our linear mapping, Eq. (\ref{eq:mapping}).
We measured the grid spacing across the extent of the CCD image. To
within an uncertainty equal to the width of the image of a grid dot,
we observed no variation in the grid spacing, confirming the absence
of nonlinear image distortion.

\subsection{Calculating Required Retardances}

\label{subsec:retardances}For a desired transformation from $\mathbf{S}=[s_{1},s_{2},s_{3}]$
to $\mathbf{S}'=[s'_{1},s'_{2},s'_{3}]$, the retardances $\zeta_{1}$
and $\zeta_{2}$ must be calculated for each desired transformation.
They are given by Sit \emph{et al.} \cite{sit:17} as, 
\begin{equation}
\zeta_{1}=\mathrm{atan2}\left([0,s_{2},s_{3}]\cdotp[0,s''{}_{2},s''{}_{3}],\textrm{sgn}(s_{2}s''{}_{3}-s_{3}s''{}_{2})\left|[0,s_{2},s_{3}]\times[0,s''{}_{2},s''{}_{3}]\right|\right)\textrm{ mod }2\pi\label{eq:zeta1}
\end{equation}
\begin{equation}
\zeta_{2}=\mathrm{atan2}\left([s''_{1},0,s''{}_{3}]\cdotp[s'_{1},0,s'_{3}],\textrm{sgn}(s'_{1}s''{}_{3}-s'_{3}s''{}_{1})\left|[s''_{1},0,s''{}_{3}]\times[s'_{1},0,s'_{3}]\right|\right)\textrm{ mod }2\pi\label{eq:zeta2}
\end{equation}
\begin{equation}
\mathbf{S}''=\left[s_{1},s'_{2},\textrm{sgn}\left(s'_{3}\right)\left(\left(s_{3}\right)^{2}+\left(s_{2}\right)^{2}-\left(s'_{2}\right)^{2}\right)^{(1/2)}\right]=[s''{}_{1},s''{}_{2},s''{}_{3}]\label{eq:sm}
\end{equation}
Where $\mathbf{S}''=\hat{R}(\mathbf{S_{1}},\zeta_{1})\mathbf{S}$
is an intermediate polarization state resulting from the first rotation
and serves as the starting point for the second rotation. The four
quadrant inverse tangent function, $-\pi\leq\mathrm{atan2(x,y)\leq\pi}$,
gives the angle from the positive $x$ axis to the point $(x,y)$.
The $\mathrm{sgn}$ function is the signum function with the convention
$\mathrm{sgn}(0)=1$. The modulo function $\mathrm{mod} 2\pi\geq0$
by convention, and $\times$ and $\cdot$ are the standard cross and
dot products, respectively. Geometrically, $\zeta_{1}$ is the positive
angle between the projections of the $\mathbf{S}$ and $\mathbf{S}''$
vectors onto the $\mathbf{S_{2}}-\mathbf{S_{3}}$ plane. Similarly,
$\zeta_{2}$ is the positive angle between the projections of the
$\mathbf{S}''$ and $\bar{\mathbf{S}}'$vectors onto the $\mathbf{S_{1}}-\mathbf{S_{3}}$
plane.



\begin{thebibliography}{10}
\bibitem{mitchell:17} K. Mitchell, N. Radwell, S. Franke-Arnold,
M. Padgett, D. Phillips, \char`\"{}Polarisation structuring of broadband
light\char`\"{}, \opex \textbf{25}(21), 25079 (2017).

\bibitem{salla:17} G. R. Salla, V. Kumar, Y. Miyamoto, R. P. Singh,
\char`\"{}Scattering of Poincaré beams: polarization speckles\char`\"{},
\opex \textbf{25}(17), 19886-19893 (2017).

\bibitem{leuchs:03} R. Dorn, S. Quabis, G. Leuchs, \char`\"{}Sharper
focus for a radially polarized light beam\char`\"{}, \pl \textbf{91}(23),
233901 (2003).

\bibitem{zhao:16} Y. Zhao, Q. Peng, C. Yi, S.G. Kong, \char`\"{}Multiband
Polarization Imaging\char`\"{}, Journal of Sensors \textbf{2016},
5985673 (2016).

\bibitem{clark:16} T. W. Clark, R. F. Offer, S. Franke-Arnold, A.
S. Arnold, and N. Radwell, \char`\"{}Comparison of beam generation
techniques using a phase only spatial light modulator\char`\"{}, Opt.
Express \textbf{24}(6), 6249 - 6264 (2016).

\bibitem{bolduc:13} E. Bolduc, N. Bent, E. Santamato, E. Karimi,
and R.W. Boyd, \char`\"{}Exact solution to simultaneous intensity
and phase encryption with a single phase-only hologram\char`\"{},
\ol \textbf{38}(18), 3546 - 3549 (2013).

\bibitem{arrizon:07} V. Arrizon, U. Ruiz, R. Carrada, and L. A. Gonzalez,
\char`\"{}Pixelated phase computer holograms for the accurate encoding
of scalar complex fields\char`\"{}, \josaa \textbf{24}(11), 3500
- 3507 (2007).

\bibitem{moreno:12} I. Moreno, J. A. Davis, T. M. Hernandez, D. M.
Cottrell, and D. Sand, \char`\"{}Complete polarization control of
light from a liquid crystal spatial light modulator\char`\"{}, \opex
\textbf{20}(1), 364 - 376 (2012).

\bibitem{chen:11} H. Chen, J. Hao, B.F. Zhang, J. Xu, J. Ding, and
H.T. Wang, \char`\"{}Generation of vector beam with space-variant
distribution of both polarization and phase\char`\"{}, \ol \textbf{36}(16),
3179 - 3181, (2011).

\bibitem{neil:02} M. A. A. Neil, F. Massoumian, R. Juskaitis, and
T. Wilson, \char`\"{}Method for the generation of arbitrary complex
vector wave fronts\char`\"{}, \ol \textbf{27}(21), 1929 - 1931 (2002).

\bibitem{maurer:07} C. Maurer, A. Jesacher, S. Furhapter, S. Bernet,
and M. Ritsch-Marte, \char`\"{}Tailoring of arbitrary optical vector
beams\char`\"{}, New Journal of Physics \textbf{9}(3), 78, (2007).

\bibitem{wang:07} X.L. Wang, J. Ding, W.J. Ni, C.S. Guo, and H.T.
Wang, \char`\"{}Generation of arbitrary vector beams with a spatial
light modulator and a common path interferometric arrangement\char`\"{},
\ol \textbf{32}(24), 3549 - 3551 (2007).

\bibitem{franke-arnold:07} S. Franke-Arnold, J. Leach, M. J. Padgett,
V. E. Lembessis, D. Ellinas, A. J. Wright, J. M. Girkin, P. Ohberg,
and A. S. Arnold, \char`\"{}Optical ferris wheel for ultracold atoms\char`\"{},
\opex \textbf{15}(14), 8619 - 8625 (2007).

\bibitem{han:15} W. Han, W. Cheng, and Q. Zhan, \char`\"{}Design
and alignment strategies of 4f systems used in the vectorial optical
field generator\char`\"{}, \ao \textbf{54}(9), 2275 (2015).

\bibitem{waller:13} E. H. Waller and G. von Freymann, \char`\"{}Independent
spatial intensity, phase and polarization distributions\char`\"{},
\opex \textbf{21}(23), 28167 (2013).

\bibitem{maluenda:13} D. Maluenda, I. Juvells, R. Martinez-Herrero,
and A. Carnicer, \char`\"{}Reconfigurable beams with arbitrary polarization
and shape distributions at a given plane\char`\"{}, \opex \textbf{21}(5),
5432 - 5439 (2013).

\bibitem{guo:14} C.S. Guo, Z.Y. Rong, and S.Z. Wang, \char`\"{}Double-channel
vector spatial light modulator for generation of arbitrary complex
vector beams\char`\"{}, \ol \textbf{39}(2), 386 (2014).

\bibitem{chen:15} H. Chen, T. Huang, J. Ding, and G. Li, \char`\"{}Independent
and simultaneous tailoring of amplitude, phase, and complete polarization
of vector beams\char`\"{}, arXiv:1510.08363 (2015).

\bibitem{davis:2000} J. A. Davis, D. E. McNamara, D. M. Cottrell,
and T. Sonehara, \char`\"{}Two-dimensional polarization encoding with
a phase-only liquid-crystal spatial light modulator\char`\"{}, \ao
\textbf{39}(10), 1549 - 1554 (2000).

\bibitem{eriksen:01} R. L. Eriksen, P. C. Mogensen, and J. Gluckstad,
\char`\"{}Elliptical polarisation encoding in two dimensions using
phase-only spatial light modulators\char`\"{}, \oc \textbf{187}(4),
325 - 336 (2001).

\bibitem{kenny:12} F. Kenny, D. Lara, O. G. Rodriguez-Herrera, and
C. Dainty, \char`\"{}Complete polarization and phase control for focus-shaping
in high-na microscopy\char`\"{}, \opex \textbf{20}(13), 14015-14029
(2012).

\bibitem{estevez:15} I. Estevez, A. Lizana, X. Zheng, A. Peinado,
C. Ramirez, J. L. Martinez, A. Marquez, I. Moreno, and J. Campos,
\char`\"{}Parallel aligned liquid crystal on silicon display based
optical setup for the generation of polarization spatial distributions\char`\"{},
\emph{Proc. SPIE} \textbf{9526}, 95261A-10 (2015).

\bibitem{zheng:15} X. Zheng, A. Lizana, A. Peinado, C. Ramirez, J.
L. Martinez, A. Marquez, I. Moreno, and J. Campos, \char`\"{}Compact
lcos - slm based polarization pattern beam generator\char`\"{}, \emph{J.
Lightwave Technol.} \textbf{33}(10), 2047 - 2055 (2015).

\bibitem{galvez:12} E. J. Galvez, S. Khadka, W. H. Schubert, and
S. Nomoto, \char`\"{}Poincare-beam patterns produced by nonseparable
superpositions of laguerre-gauss and polarization modes of light\char`\"{},
\ao \textbf{51}(15), 2925 - 2934 (2012).

\bibitem{sit:17} A. Sit, L. Giner, E. Karimi, and J.S. Lundeen, \char`\"{}General
Lossless Spatial Polarization Transformations\char`\"{}, \emph{Journal
of Optics} \textbf{19}(9), 094003 (2017).

\bibitem{wolf:99} M. Born, E. Wolf, \textit{Principles of Optics,
Seventh Ed.}, (Cambridge University Press, 1999).

\bibitem{josza:94} R. Josza,\char`\"{}Fidelity for Mixed Quantum
States\char`\"{}, \jmo \textbf{41}(12), 2315-2323 (1994).

\bibitem{moreno:14} J. Martinez, I. Moreno, M. Sanchez-Lopez, A.
Vargas, P. Garcia-Martinez, \char`\"{}Analysis of multiple internal
reflections in a parallel aligned liquid crystal on silicon SLM\char`\"{},
Opt. Express \textbf{22}(21), 25866-25879 (2014).

\bibitem{goldstein:03} D. Goldstein, \textit{Polarized Light} (Marcel
Dekker, Inc., 2003), Chap. 27

\bibitem{chuang:00} M. Nielson, I. Chuang, \textit{Quantum Computation
and Quantum Information} (Cambridge University Press, 2000).

\bibitem{caves:04} J.M. Renes, R. Blume-Kohout, A.J. Scott, C.M.
Caves, \char`\"{}Symmetric informationally complete quantum measurements\char`\"{},
J. Math. Phys., \textbf{45}, 2171-2180 (2004).

\bibitem{chigrinov:12} H. Wu, W. Hu, H-C. Hu, X. Lin, G. Zhu, J-W.
Choi, V. Chigrinov, Y-Q Lu, \char`\"{}Arbitrary photo-patterning in
liquid crystal alignments using DMD based lithography system\char`\"{},
\opex \textbf{20}(15), 16684-16689 (2012). 
\end{thebibliography}
\end{document}